# Decreasing water budget of the Australian continent from Grace satellite gravity data

C. O'Neill[1*] & S. Chandler-Ho[1]

[1]Planetary Research Centre, Macquarie University, Sydney, NSW, 2109, Australia.
*Corresponding author. Email: craig.oneill@mq.edu.au.

## Abstract

**Increasing aridification of continental areas due to global climate change has impacted freshwater availability, particularly in extremely dry landmasses, such as Australia. Multiple demands on water resources require integrated basin management approaches, necessitating knowledge of total water storage, and changes in water mass. Such monitoring is not practical at continental scales using traditional methods. Satellite gravity has proven successful at documenting changes in total water mass at regional scales, and here we use data from the Grace and Grace-FO missions, spanning 2002 - 2020, to track regional water budget trends in Australia most heavily utilised basin systems, including the Murray-Darling Basin. The period of analysis covers the Millennium drought (2002-2009) and 2010-11 heavy flooding events, which contribute significant signal variability. However our extended datasets demonstrate a negative trend in the geoid anomaly over the Murray-Darling Basin of -1.5mm, equivalent to a water loss rate of -0.91 Gt yr$^{-1}$. With the exception of northern Australia, similar scale geoid declines are observed in most Australian basin systems analysed - implying declining total water storage. Long-term declines in water availability require concerted management plans, balancing the requirements of agriculture and industry, with domestic use, traditional owners, and healthy freshwater ecosystems.**

## Introduction

Global climate change has seen the increasing aridity over continental areas (1), putting a strain of existing water resources and agriculture (2). Australia is the driest inhabited continent (3), and is particularly vulnerable to the consequences of increasing aridity, including devastating bushfires (3), prolonged extreme drought events (2), agricultural failure due to extreme weather events (2), destruction of native wildlife and environment degradation (4), and access to fresh drinking water (5).

Whilst critical basin areas in Australia are subject to hydrological monitoring, trends in basin-level water budgets are problematic to calculate, due to uncertainties in i) groundwater resources, which at a continent scale are largely unknown, ii) soil moisture variation, and iii) river flow budgets, which may have overestimated historical flows, and are uncertain due to unregistered usage of river water and flood water reclamation (6). To calculate the basin-scale risk of increasing water aridity across Australia, there is a need to develop a method for monitoring continental water budget.

The GRACE (Gravity Recovery and Climate Experiment) mission operated from 2002-2017, and used a constellation of two satellites to derive gravity field models from their displacement, and satellite tracking (7). It was developed to monitor the global gravity field through time, allowing an assessment of the changing distribution of Earth's water masses (8,9). It is sensitive to seasonable wet-dry season variation in equatorial areas (eg. the Amazon, and equatorial Africa), and has been used to document groundwater depletion in southern California and northern India (10). Its success spurred the development of the Grace Follow-On mission (Grace-FO), which launched in 2018 and is ongoing.

Previous assessment of water budget trends for Australia (10), based on 2002-2016 Grace data, noted that the signal was dominated in Eastern Australia by i) an extreme drought between 2001-2009, followed by ii) extreme flooding events in 2011-2012. The latter was caused by an extreme La Niña event in those years, which was probably exacerbated by global warming (11). As the timescale of the analysis (2002-2016) was of a similar scale to the La-Niña - El-Niño variability (the El Niño - Southern Oscillation, or ENSO), these events precluded any significant trend. In NW Australia, the beginning of the GRACE mission coincided with the end of a significant period of downfall, and, again the short time period prevented the identification of secular trends (10).

Since the analysis of (10), further gravity data from the GRACE-FO has become available, which permits the extension of this data over several La Niña cycles, allowing the identification of trends beyond the 2010-2011 La Niña flooding event. In our analysis below we focus on the Murray-Darling basin (MDB, Figure 1) - the drainage basin of Australia's largest river system, the Murray Darling. The region covers Australia's most extensively irrigated areas (12), and is responsible for nearly 40% of Australia's total value of agricultural production (12), including water-demanding crops such as cotton, rice, and also almond trees. Despite the development of a management plan for the basin (13), continued flow restrictions in the river systems have seen mass-fish die-offs (14), and placed significant stress on freshwater ecosystems, which rely on regular floodwaters for reproduction (14). To put the MDB in context, we will also assess the broader continent-scale trends in water mass.

## Results

Variation in the regional geoid signal over the Murray Darling Basin is shown in Figure 2. The Grace and Grace-FO was gridded and plotted within the MDB, and the same scale adopted between Figures to illustrate spatial and temporal trends. The snapshots are from near the beginning of the Grace mission (2003), and the months represent both the hottest and generally period of the year (January, in the Australian summer), and the wettest (July/August). The spatial variability is generally due to run-off of seasonally sporadic rains. The major river systems drain to the southwest, and the region is generally more arid in the north and west. The data from 2003 and 2016 are from the original Grace mission - encapsulating the time-trend over the course of the mission. The 2020 data are from Grace-FO.

Increased geoid signal in Figure 2 generally corresponds to increase local mass due to the presence of water. It is not sensitive to the location of the water though, and the signal has contributions from surface water (river, floodwaters), soil moisture and groundwater. Within 2003, the geoid signature of the MDB regularly exceeded the mean field average - consistent with increased water mass (relative to the 2002-2016 average), even in the summer months in the southern MDB. However, by 2016 the average geoid anomaly for both months displayed is negative, consistent with significant bulk water loss. The latest snapshots from 2020 consistently saturate the lower end of the scale.

To quantify the basin-scale effects of water-mass loss, we have plotted a monthly time-series of the basin-average geoid in Figure 3a. The raw data is shown in grey, the filtered is in blue, and the overall linear-regression trend is red.  The break in the data between 2016-2018 represents the time period between the cessation of the Grace mission, and of Grace-FO becoming operational.  Within the data, the seasonal cycle of dry, hot summers (geoid lows) and wetter winters (geoid highs) is apparent, due to excess water mass during the winter months.

The effect of the 2001-2009 Millennium drought in this region is observable in the slow trend towards lower geoid readings in this interval - however, the extreme flooding events of 2010-2011 overwhelm this signal, temporarily increasing the average geoid signal by 3mm. These signals are also seen in rainfall data - we have plotted rainfall data from Menindee (western NSW) as a comparison (Figure 3c), with the 2010 floods extremely obvious in this timeseries. These weather patterns, which have pseudo-decadal variability, obscure any long-term trend in the data from the original Grace mission, as previously shown (10).

With the Grace-FO mission, however, the overall 18 year trend in the data becomes apparent, with the average baseline geoid dropping 1.5mm over this period (red line, Figure 3a). This is less than the variability seen due to (for instance) La Niña events, but it is coherent. The average for Grace-FO is even less than this trend, with the geoid signal during the summer 2020 extreme bushfire season in eastern Australia - a product of extremely dry arid conditions - was the lowest monthly average observed since the missions began in 2002.

At a continent-scale, Australia's climate systems are affected not just by the Pacific La-Niña/El Niño system, but the Indian-Ocean dipole, and the Southern Annual Mode (17), as well as wet-dry system variability in northern Australia. These systems operate differently in different parts of Australia, and impact long-term regional climate predictions (17). To assess the regional variability over continental-Australia, we have constructed time-series of the Grace and Grace-FO geoid data again in Figure 4, for a number of locales, by interpolating the raw geoid grid data to a defined point.

Across northern Australia, there is little observed trend in the geoid - due to strong rains in the wet season - but it is worth noting there is no observed increase in the water budget, either. NW Australia, near the Kimberley region, was identified as having a strong negative water storage trend in (10). We observe a similar, but moderate, negative trend in the geoid.

The Pilbara and SW Australia (mining and agricultural hubs, respectively) both show strong negative geoid trends (>4mm from 2002-2020), which are well above the seasonal and decadal variability.  Central Australia and Adelaide both show decreasing trends, with the caveat that the Adelaide sample point is near the coast and may be influenced by ocean signal.

Western Queensland, the NSW Tablelands, and the Australian High-Country (Victoria) cover a range of environments within the eastern states, from semi-arid farmland (W. Queensland), to forested hinterlands (NSW Tablelands), to Alpine (Highlands). The latter is unique in being one of the few mainland areas to receive regular snowfall, and is proximal to some of Australia's most significant hydroelectric projects (Snowy Hydro 1.0 and 2.0). The signals in all three regions are strongly perturbed by the Millenium drought and 2010-11 flooding events, but show a consistent downward trend in the geoid, indicating long-term water mass loss.

Lastly, western Tasmania - a region dominated by Southern Ocean weather, and with high annual rainfalls, also shows a very clear negative trend in geoid and water mass. The point is near the coast, and thus possibly has an ocean signal contribution, but the trend exceeds seasonal variability significantly.

## Discussion and Conclusions

The long-term geoid trends observed for most parts of the Australian continent suggest a rather dire projected water budget, in what is already an extremely arid environment. With the exception of northern Australia, Grace satellite gravity demonstrates significant long-term negative trends in the regional geoid, consistent with large-scale water-mass loss from continental Australia. The Grace and Grace-FO signals are sensitive to mass changes, and thus by themselves do not differentiate between surface water, soil moisture, and groundwater resources.

Groundwater estimates are particularly problematic, as they represent one of the largest unknowns in the water budget at regional scales. Locally, aquifers are typically monitored using bores, and, together with detailed models of recharge, discharge, and local geology, groundwater models may be constructed. This level of information is not available at continental scales, and may also exclude deep aquifers. In this context geophysical approaches - such as satellite gravity - are invaluable, in conjunction with surface hydrology knowledge, in constraining total water budget changes.

Groundwater and surface water are course coupled by recharge and discharge, and changes in surface water supply can have a delayed response on groundwater budget. Behavioural responses to surface water variations, such as drought, can also lead to extensive extraction of groundwater. In the case of southern California, increased groundwater usage during recent droughts has lowered groundwater levels, and the dewatering has irrevocably reduced aquifer capacity due to compaction (10). Estimates of the total stored water loss are of the order -4.2 ± 0.4 Gt yr$^{-1}$ (10).

For the Murray-Darling Basin, the linear trend equates to a loss of an equivalent water layer of 1.57cm from 2002-2020. Over the area of the basin, this equates to a loss of 16.59 Gt, or around -0.91 Gt yr$^{-1}$, around 22% the loss rate of southern California, over a much larger area. There is a large variability in the total supply of water to the MDB over decadal timescales, but estimates are of the order ~531 Gt yr$^{-1}$ (12). Of this, over 94% evaporates or transpires, 4% is run-off, and 2% - ~10.6 Gt  yr$^{-1}$ - drains into ground (12). Thus a yearly loss of -0.91 Gt is a significant fraction (8.5%) of estimated water delivered into the ground system.

The observed downward geoid, and equivalent water height, trend in the Murray-Darling Basin put increasing demands on agriculture, and water-intensive crops (such as cotton or rice) and open irrigation may become unviable. Similar issues are raised for western Australia. Southwestern Western Australia is the country's largest wheat producing region, and an agricultural hub (18), and the Pilbara region is an immense iron-ore producer (19). Decreasing water budget creates significant challenges for these arid, but economically significant, portions of Australia. Similarly, aridification of the Adelaide area puts stress not just on one of Australia's largest wine-growing regions (20). Adelaide's primary freshwater source is the Murray-Darling basin (5), and increasing local aridity places additional demands on an already overtaxed river system.

Monitoring global water resources is critical in the context of global climate change, particularly in arid environments - such as much of Australia - where water availability is already marginal, and any net system loss requires considered management of competing demands, such as agriculture, mining, manufacturing, and domestic use. The Murray-Darling Basin is arguably Australia's most important agricultural region, and water resources are under considerable stress (2,5,6). Our analysis of geoid variations from the Grace satellite gravity missions suggest a significant negative trend of -1.5mm over 2002-2020, equating to a yearly loss of -0.91 Gt - a significant fraction of water delivered to soil and groundwater systems.  With the exception of Northern Australia, most other regions explored here show declining trends in water mass, although the Millennium drought and 2010-11 La Niña flooding events contribute significant signal variability. Increased water stress in Australia is likely to have flow-on impacts, from freshwater ecosystems, to increased bushfire risk (3), and suggest arid regions globally will need a concerted effort in water resource management and monitoring in a changing global climate.

## Methods

We utilise the processed RL05 geoid and equivalent water height (EWH) data from the Grace and Grace-FO missions, from CNES-GRGS (15). The data are gridded from spherical harmonic coefficients up to degree 90, with low degrees supplemented by SLR observations (15). The geoid anomalies are calculated relative to an average gravity model [EIGEN-GRGS.RL04.MEAN-FIELD]. We have extracted the data within a Murray-Darling shapefile, and regional trends within the basin, as well as outside it, to eliminate peripheral noise sources (eg. ocean temperature variability - (16)). The process is shown in Figure 1, which illustrates the regional resolution of the Grace data over the Australian continent (Figure 1a), and a masking of data within the Murray-Darling shapefile (Figure 1b, see S1).

We average the data within the Murray-Darling region, and the raw data is shown in Figure S1. We have also plotted the raw data for the northern Murray-Darling basin (latitude > -30S). These two time-series show minor regional variations, but are broadly consistent, demonstrating that ocean signal is not significant. We then apply a smoothing filter to our data, and calculate the long term across both the Grace and GRACE-FO mission, from 2002-2020 (Figure 3).

## Acknowledgements

CO acknowledges funding from the Planetary Research Centre, Macquarie University, and the ARC (DP210102196). Colormaps utilised scientific color schemes of (21).

## Data availability

Data is available to download from the CNES-GRG website, https://grace.obs-mip.fr/, and the wonderful Grace Plotter site: http://thegraceplotter.com/.

## Code availability

The Jupyter notebook used to generate the plots and analysis here is available via the online repository Zenodo.org (TBC).

## Author contributions

CO instigated and developed the study, developed code and analysis procedures, created final figures, and wrote the manuscript. SC-H undertook processing of data, and visualisation.

## Competing interests

The authors have no competing interests.

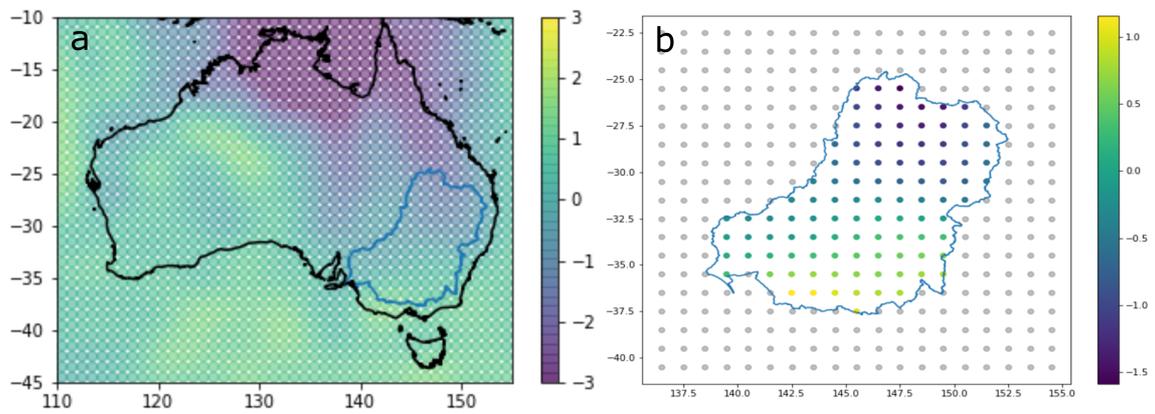

Figure 1. a) Graphical representation of the RL05 CNES-GRGS gridded geoid anomaly Grace product, at original grid spacing, together with coastlines (black), and the extent of the Murray-Darling Basin (blue). Geoid measured in mm. b) Grace geoid anomaly data within the extent of the Murray-Darling Basin (coloured, geoid in mm) used in the regional analysis. Monochrome points falls outside the basin and are excluded.

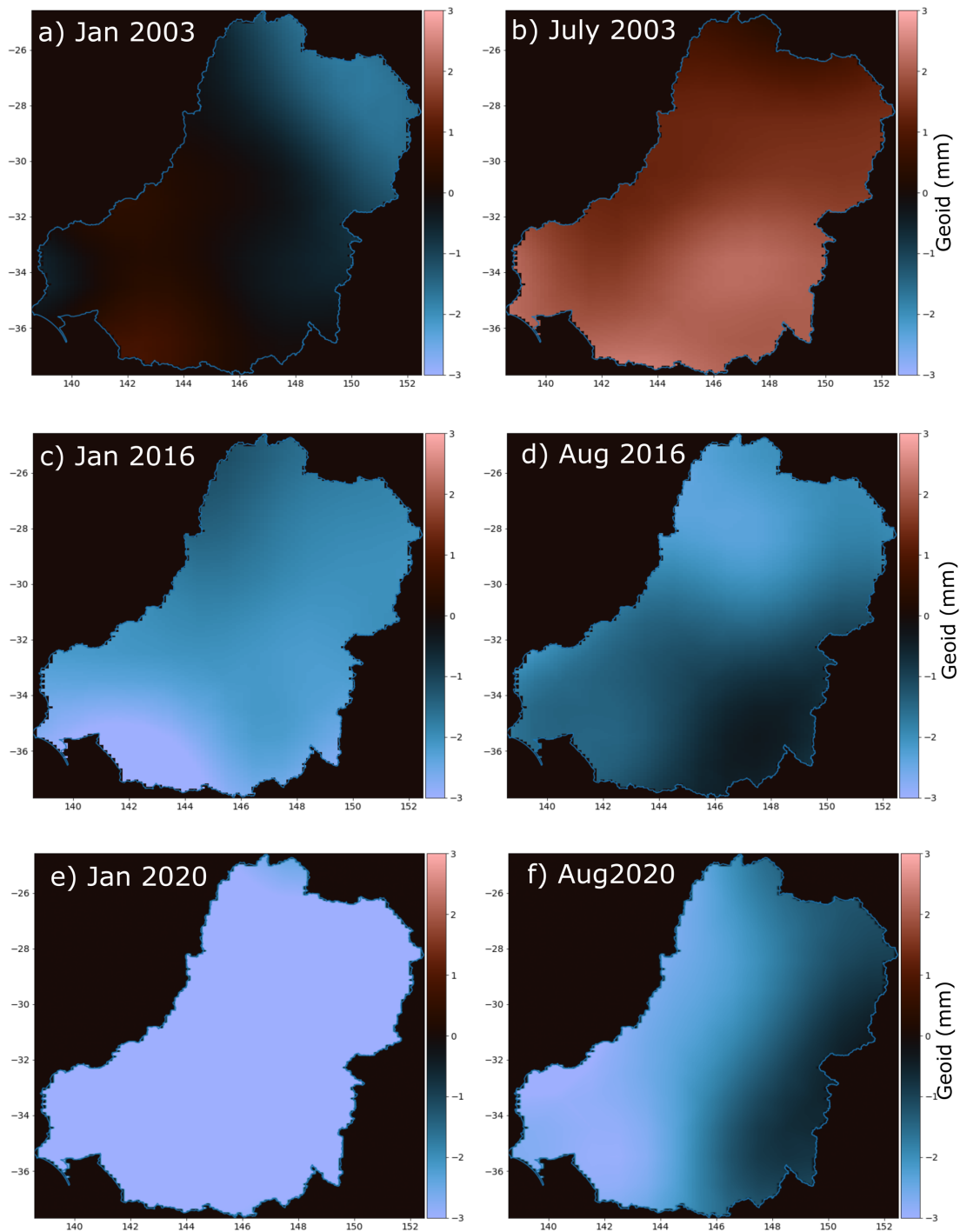

Figure 2. Geoid anomaly within the Murray-Darling Basin, Australia, from 2003-2020. Snapshots a-f are for different times (labelled), showing the annual variability between dry summer weather (January), and wet winter weather (July-August; excess mass from winter rainwater contributes to positive geoid signals). A long-term trend towards lower geoid values is apparent across the basin.

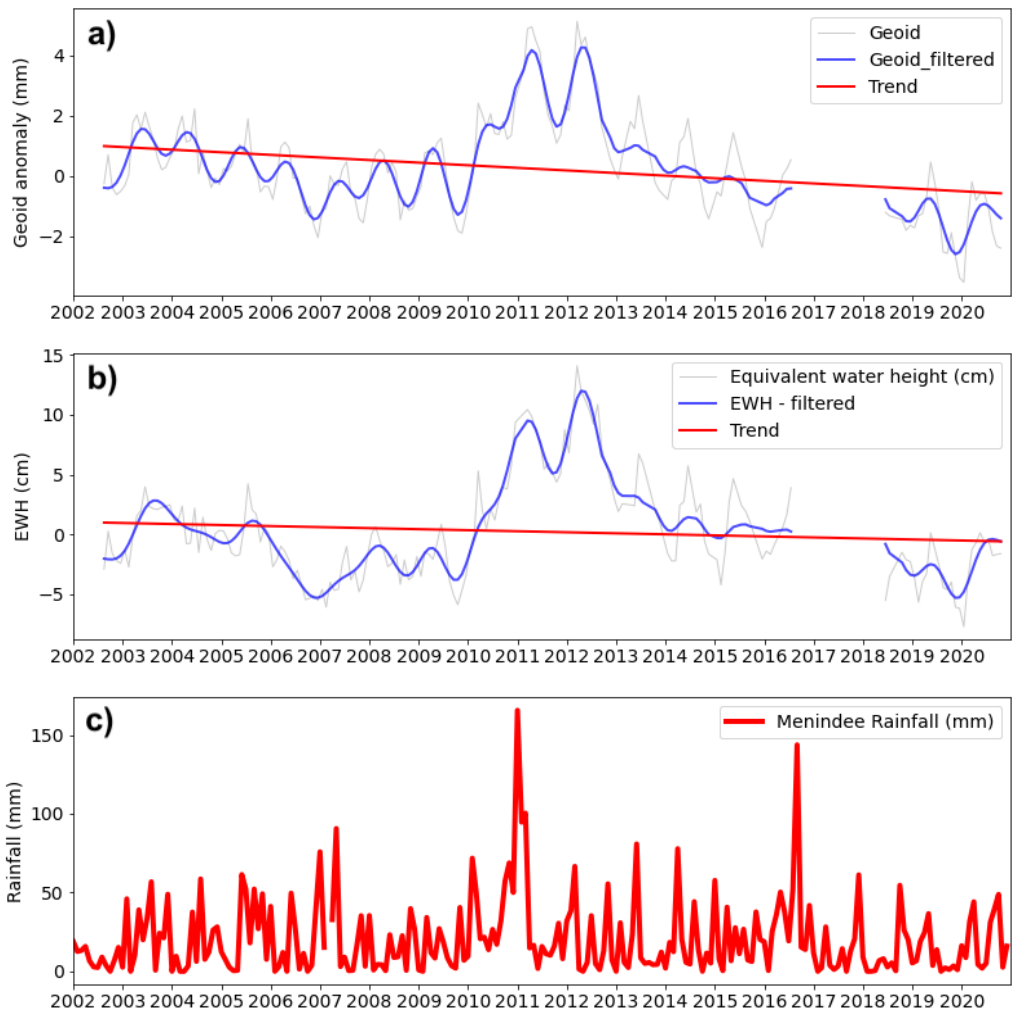

Figure 3. a) Monthly geoid anomaly for the Murray-Darling basin (MDB). Raw data is grey, filtered data blue, and the linear fitted trend is shown in red. The break in the data between 2016-2018 represents the time-period between the cessation of Grace, and the launch and operations of Grace-FO. b) Equivalent water layer height, calculated from the mass changes implied by the geoid signal, for the Murray Darling. Grey is raw monthly mean data for the MDB, blue is the filtered signal, and red the linear trend. c) Documented rainfall at Menindee (32.37S, 142.40E, within the Murray-Darling Basin, over the time of the Grace dataset.

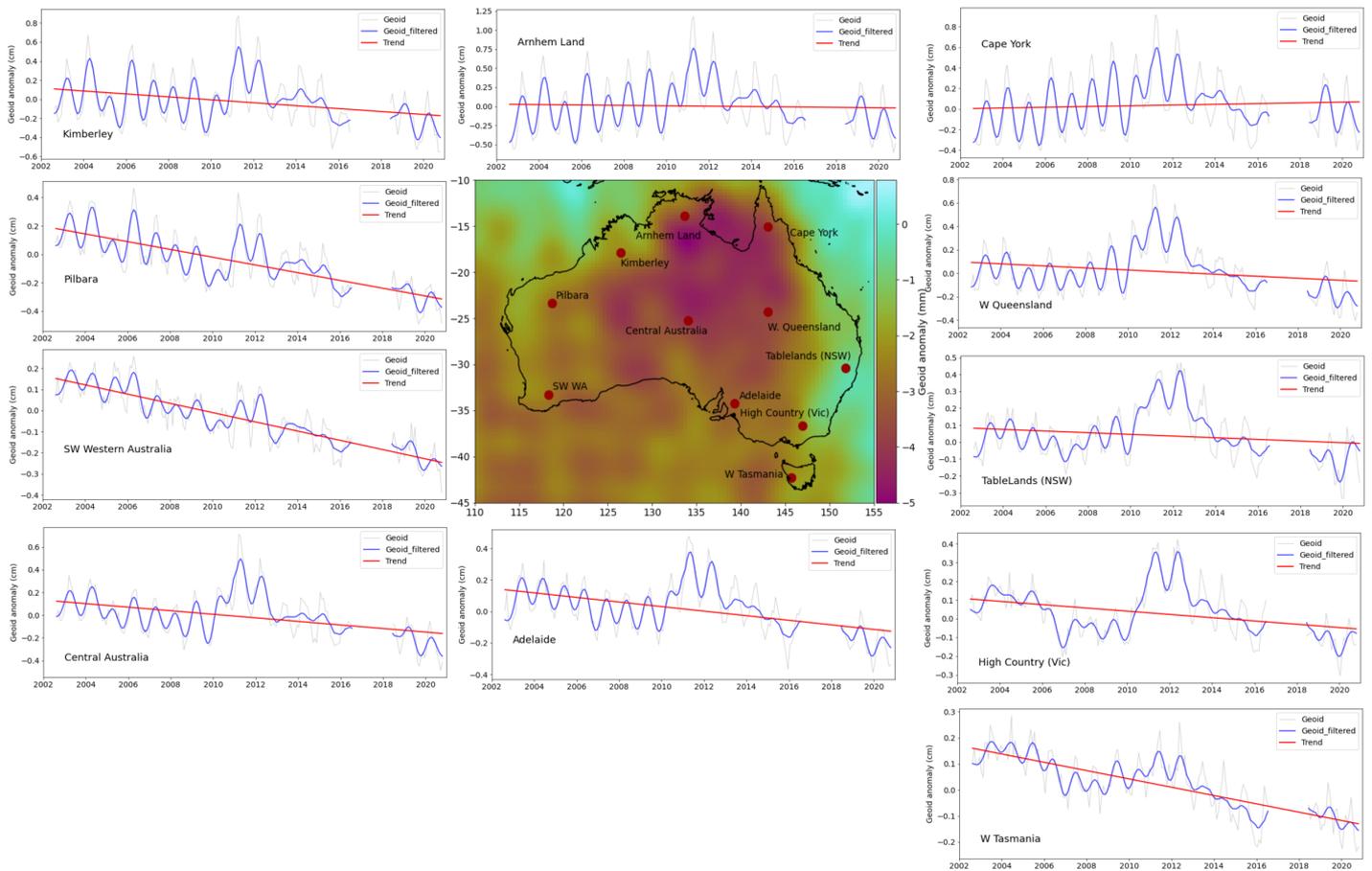

Figure 4. Grace geoid anomaly (mm), calculated by interpolation at the points shown in the map inset. Grey is raw data, blue filtered, and red is the linear trend over the period of data availability. Negative geoid trends imply diminished water mass. Colours in the map inset show geoid anomaly for August, 2020.

# Supplementary Information

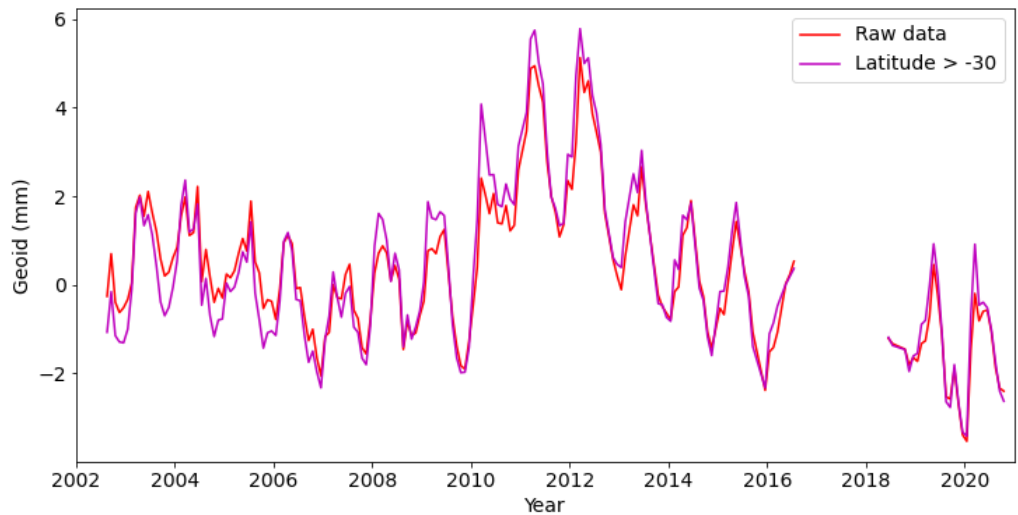

Supplementary Figure S1. Timeseries of raw mean geoid anomaly data from the Murray-Darling Basin, from the Grace CNES-GRGS geoid solutions. Red shows the mean value for the entire basin, magenta for the northern basin above latitude -30, to assess noise contributions in the basin from ocean and regional variations.